\documentclass[pra,aps,showpacs,groupedaddress,superscriptaddress,twocolumn,toc=flat]{revtex4-1}

\usepackage[utf8x]{inputenc}
\usepackage{color}
\usepackage{bbm} 

\usepackage{amsfonts,amsmath,amssymb,stmaryrd}

\usepackage{graphicx}
\usepackage{subfigure}  
\usepackage{bbm} 
\usepackage{hyperref}
\usepackage{epsfig}
\usepackage{mathrsfs}
\usepackage{verbatim}
\usepackage{centernot}
\usepackage{ulem}

\renewcommand{\l}{\left(}
\renewcommand{\r}{\right)}

\newcommand{\bra}[1]{\langle#1|}
\newcommand{\ket}[1]{|#1\rangle}

\renewcommand{\H}{\hat{\mathcal{H}}}

\renewcommand{\a}{\hat{a}}

\newcommand{\ad}{\hat{a}^\dagger}

\newcommand{\hc}{\text{h.c.}}

\usepackage{array}

\usepackage{cancel,ifthen}
\newcommand{\cmnt}[2][NoInPuT]{\ifthenelse{\equal{#1}{NoInPuT}}{}{{\color{red}\sout{#1}}} {\color{blue} #2}}

\usepackage{bm}	
\renewcommand{\vec}[1]{\bm{#1}}

\begin{document}
\normalem	

\title{2D spin-orbit coupling for ultracold atoms in optical lattices}

\author{Fabian Grusdt}
\affiliation{Department of Physics, Harvard University, Cambridge, Massachusetts 02138, USA}

\author{Tracy Li}
\affiliation{Fakult\"at f\"ur Physik, Ludwig-Maximilians-Universit\"at M\"unchen, Schellingstrasse 4, 80799 Munich, Germany}
\affiliation{Max-Planck-Institut f\"ur Quantenoptik, Hans-Kopfermann-Str. 1, 85748 Garching, Germany}
\affiliation{Department of Physics, Stanford University, Stanford, California 94305, USA}

\author{Immanuel Bloch}
\affiliation{Fakult\"at f\"ur Physik, Ludwig-Maximilians-Universit\"at M\"unchen, Schellingstrasse 4, 80799 Munich, Germany}
\affiliation{Max-Planck-Institut f\"ur Quantenoptik, Hans-Kopfermann-Str. 1, 85748 Garching, Germany}

\author{Eugene Demler}
\affiliation{Department of Physics, Harvard University, Cambridge, Massachusetts 02138, USA}


\date{\today}

\begin{abstract}
Spin-orbit coupling (SOC) is at the heart of many exotic band-structures and can give rise to many-body states with topological order. Here we present a general scheme based on a combination of microwave driving and lattice shaking for the realization of time-reversal invariant 2D SOC with ultracold atoms in systems with inversion symmetry. We show that the strengths of Rashba and Dresselhaus SOC can be independently tuned in a spin-dependent square lattice. More generally, our method can be used to open gaps between different spin states without breaking time-reversal symmetry. We demonstrate that this allows for the realization of topological insulators in the presence of SOC, which is closely related to the Kane-Mele model.
\end{abstract}

\maketitle

\emph{Introduction.--}
The band theory of solids developed in the 1920's is a cornerstone of modern physics and has been essential for understanding the properties of a wide range of materials. Even today, the calculation of full band structures from microscopic models is an active field, challenged by complications arising from the interplay of single-particle physics with many-body interaction effects. Quite remarkably, {it was recently realized that even} the geometric structure of single-particle Bloch wavefunctions can be sufficiently involved to give rise to distinct topological effects. This new class of states, termed topological insulators \cite{Kane2005,Bernevig2006a,Bernevig2006,Fu2007,Koenig2007,Hsieh2008,Hasan2010,Qi2011}, possess intriguing transport properties due to, for example, the presence of topologically protected edge states, where spin and momentum are correlated. A key requirement for realizing topological insulators is the presence of time-reversal (TR) invariant spin-orbit coupling (SOC) \cite{Kane2005,Bernevig2006a,Bernevig2006}.

Beyond the single-particle picture, interaction effects can significantly enrich the physics of band insulators. For example, in the fractional quantum Hall effect \cite{Tsui1982}, the interplay between topologically non-trivial Bloch bands \cite{Haldane1988} and strongly-correlated electrons gives rise to quasiparticles with fractional charges \cite{Laughlin1983} and anyonic braiding statistics \cite{AROVAS1984}. Along the same lines, it has recently been suggested that interacting bosons subject to SOC can fermionize in two dimensions and form exotic many-body states \cite{Sedrakyan2012}. However, the physics of strongly interacting particles in the presence of large SOC has not yet been explored in depth, partly due to the relatively weak SOC achievable in solids. 

Ultracold quantum gases provide a promising alternative platform for experimentally studying this situation \cite{Galitski2013}. In these systems, strong interactions are routinely obtained by increasing the trap depth or utilizing Feshbach resonances~\cite{Bloch2008}. A common route to SOC in homogenous systems is to generate synthetic gauge fields by engineering the geometric phases acquired by neutral atoms dressed by Raman lasers~\cite{Ruseckas2005,Dalibard2011, Juzeliunas2010,Campbell2011}. Using this technique, SOC was implemented in both 1D~\cite{Lin2011,Wang2012,Cheuk2012} and 2D homogenous systems~\cite{Huang2016} in the absence of an optical lattice.

Experimentally realizing SOC in optical lattices has proven to be more challenging. Theoretical proposals consist of generalizations of laser-assisted tunneling schemes \cite{Jaksch2003} to generate spin-dependent hoppings \cite{Osterloh2005,Goldman2010,Goldman2013JphysB}, lattice shaking approaches~\cite{Struck2014}, adiabatic schemes \cite{Dudarev2004} related to optical flux lattices \cite{Cooper2011} and an intriguing use of optical lattice clocks \cite{Wall2016}. Thus far, SOC has been experimentally realized in 1D optical lattices by using different motional states for the two spin degrees of freedom \cite{Hugel2013, Atala2014,LiKetterle2016}. In 2D optical lattices, a version of SOC which breaks TR symmetry was recently realized using an optical Raman scheme \cite{Liu2014,Wu2016}.

\begin{figure}[b!]
\centering
\epsfig{file=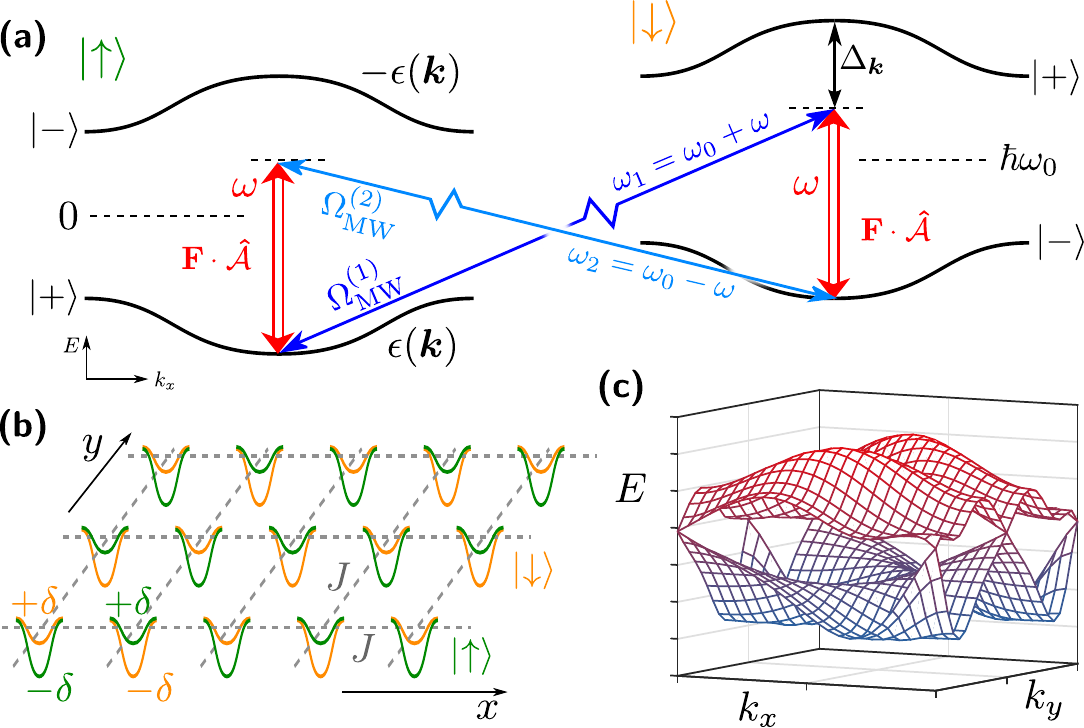, width=0.45\textwidth}
\caption{SOC in optical lattices: (a) Starting with two pseudospins $\ket{\! \! \uparrow}$ and $\ket{\! \! \downarrow}$ with band structures related by reversing their order and offsetting them by energy $\hbar \omega_0$, tunable SOC is realized for states $\ket{ \! \! \Uparrow} = \ket{ \! \! \uparrow,+}$ and $\ket{ \! \! \Downarrow} = \ket{ \! \! \downarrow,-}$ by a combination of microwave driving and lattice shaking. The latter depends on the Berry connection $\vec{\mathcal{\hat{A}}}$ (see Eq.~\eqref{eq:BerryCnctn}) and can be implemented by an oscillatory force $\vec{F}(t)$ with frequency $\omega$. The scheme can be realized, e.g., in an inversion-symmetric, spin-dependent square lattice (b), and  lifts the spin degeneracy in the band structure up to Kramer's degeneracies (c).}
\label{IntroSOC}
\end{figure}

Here, we introduce a general scheme for realizing  2D SOC in optical lattices of Rashba \cite{Bychkov1984} or Dresselhaus \cite{Dresselhaus1955} type, described by the Hamiltonians
\begin{equation}
\H_{\rm R,D} = \alpha_{\rm R,D} \l \hat{\sigma}^x k_y \mp \hat{\sigma}^y k_x \r.
\end{equation}
Our method relies on a combination of microwave (MW) drives to change the spin states of the atoms and lattice shaking to couple the spins with the atomic motion, similar to the scheme suggested by Struck et al.~\cite{Struck2014} for 1D SOC and to experimental implementations of artificial gauge fields~\cite{Kolovsky2011a,Struck2012,Aidelsburger2013,Miyake2013,Jotzu2014}. If the optical lattice is inversion symmetric, the resulting SOC can be TR invariant. We show that this allows for the realization of TR invariant topological insulators with non-trivial spin textures in the presence of spin-dependent synthetic magnetic fields \cite{Aidelsburger2013}.

\emph{General procedure.--}
We consider two independent (pseudo-) spin states $\ket{\!\!\uparrow}$ and $\ket{\!\!\downarrow}$ in 2D, both of which are subject to the same dispersion relation $\epsilon(\vec{k})$. We assume that the corresponding Hamiltonian is invariant under the anti-unitary TR operator $\hat{\theta}$,
\begin{equation}
\H(\vec{k}) = \hat{\theta}^\dagger \H(-\vec{k}) \hat{\theta},
\end{equation}
which implies that $\epsilon(-\vec{k}) = \epsilon(\vec{k})$ is symmetric.

In Fig.~\ref{IntroSOC} (a) we show the specific model which we use to derive SOC in a square lattice, where the energy of identical Bloch wavefunctions is reversed for different spin states. In this case $\hat{\theta} = K i \hat{\sigma}^y \otimes \hat{\tau}^x$ consists of a spin-flip ($i \hat{\sigma}^y$) and a simultaneous change of the Bloch band described by $\hat{\tau}^x = \ket{+}\bra{-} + \hc$, where $K$ denotes complex conjugation.

Our goal is to engineer SOC, which mixes the two pseudo-spin states $\ket{ \! \uparrow}, \ket{\! \downarrow}$ while preserving TR. That is, the resulting Hamiltonian should  be TR invariant and should not commute with $\hat{\sigma}_z$. In general, we need to engineer terms of the form 
\begin{equation}
\H_{\rm SOC}(\vec{k}) = \vec{f}(\vec{k}) \cdot \hat{\vec{\sigma}},
\label{eq:SOCterms}
\end{equation}
which are TR invariant if $\vec{f}(\vec{k})$ is an odd function of $\vec{k}$, i.e. $\vec{f}(\vec{k}) = - \vec{f}(-\vec{k})$. The lattice versions of Rashba and Dresselhaus SOC are obtained by choosing 
\begin{equation}
\vec{f}_{\rm R,D}(\vec{k}) = \alpha_{\rm R, D} (\sin k_y, \mp \sin k_x)^T.
\label{eq:fRD}
\end{equation} 

To engineer the terms in Eq.\eqref{eq:SOCterms}, we consider a second order Raman transition as shown in Fig.~\ref{IntroSOC} (a), which resonantly couples $\ket{\! \! \Uparrow} = \ket{\! \! \uparrow, +}$ to $\ket{\! \! \Downarrow} = \ket{\! \! \downarrow, -}$ through the virtually excited states $\ket{\! \! \downarrow, +}$ and $\ket{\! \! \uparrow, -}$.  While the specific model considered so far guarantees an ideal Franck-Condon overlap for this process, it is not a strictly necessary condition for our protocol to work. 

Spin flips between $\ket{\! \! \uparrow}$ and $\ket{\! \!\downarrow}$, with energy cost $\hbar \omega_0$, are achieved by two direct MW transitions $j=1,2$ with Rabi frequencies $\Omega_{\rm MW}^{(j)}$. The second leg of the Raman transition between bands $\ket{+}$ and $\ket{-}$ of the same spin state is realized by near-resonant lattice shaking with frequency $\omega$. This coupling is described by \cite{Grusdt2014Z2,Dunlap1986,Li2016}
\begin{equation}
\H_{F}(\vec{k},t) = \vec{F}(t) \cdot \vec{\mathcal{\hat{A}}}(\vec{k}),
\end{equation}
where $\vec{F}(t) = (F_x \cos (\omega t), F_y \cos(\omega t + \phi_F))^T$ is an oscillatory force, which we assume to be weak: $Fa \ll \hbar \omega$, where $a$ is the lattice constant. Without loss of generality, we fix the overall phase of lattice shaking. The elements of the $U(2)$ Berry connection are 
\begin{equation}
\vec{\mathcal{A}}^{\mu,\nu}(\vec{k}) = \bra{u_\mu(\vec{k})} i \vec{\nabla}_{\vec{k}} \ket{u_\nu(\vec{k})},
\label{eq:BerryCnctn}
\end{equation}
with $\mu,\nu = \pm$ and where $\ket{u_\mu(\vec{k})}$ denotes the cell-periodic Bloch wavefunction corresponding to the state $\ket{\mu}$ at quasimomentum $\vec{k}$.

For two-photon resonance, the MW frequencies are $\omega_{1,2} = \omega_0 \pm \omega$. Assuming $\omega_0 \gg \omega$ we obtain an effective Rabi coupling between $\ket{\! \! \Uparrow}$ and $\ket{\! \! \Downarrow}$ of the form
\begin{equation}
\H_{\rm SOC}(\vec{k}) \propto \sum_j \frac{\hbar \Omega_{\rm MW}^{(j)}}{\Delta_{\vec{k}}} \vec{F} \cdot \vec{\mathcal{A}}^{-,+}(\vec{k})  ~ \ket{\! \Downarrow} \bra{\Uparrow \!} + \hc ~ .
\label{eq:EffRabiCouplings}
\end{equation}
The $\vec{k}$-dependent detuning from the intermediate state is given by $\Delta_{\vec{k}} = 2 \epsilon(\vec{k}) + \hbar \omega$ if the band structure is particle-hole symmetric, as in Fig.~\ref{IntroSOC}a. Note that Eq.\eqref{eq:EffRabiCouplings} is in the form of Eq.\eqref{eq:SOCterms}, where $\hat{\vec{\sigma}}$ is defined in the basis of $\ket{ \! \! \Uparrow}$, $\ket{\! \! \Downarrow}$ and $\vec{f}(\vec{k})\equiv\frac{ \hbar \Omega_{\rm MW}^{(j)}}{\Delta_{\vec{k}}} \vec{F}\cdot \vec{\mathcal{A}}^{-,+}(\vec{k})$.

Eq. \eqref{eq:EffRabiCouplings} is TR invariant and thus realizes synthetic SOC if $\vec{\mathcal{A}}^{-,+}(\vec{k}) = - \vec{\mathcal{A}}^{-,+}(-\vec{k})$. To guarantee this symmetry, we consider systems which are invariant under spatial inversion, $\hat{P} \H(\vec{k}) \hat{P} = \H(-\vec{k})$. Consequently, cell-periodic Bloch states $\ket{u_\mu(\pm \vec{k})}$ are related by
\begin{equation}
\hat{P} \ket{u_{\mu}(\vec{k})} = e^{i \chi_\mu(\vec{k})} \ket{u_\mu(-\vec{k})},
\label{eq:relationInversion}
\end{equation}
where $\chi_\mu(\vec{k})$ is determined by the gauge choice. It follows that
\begin{equation}
\vec{\mathcal{A}}^{\mu,\nu}(- \vec{k}) = - e^{i \l \chi_\mu(\vec{k})- \chi_\nu(\vec{k})\r} \vec{\mathcal{A}}^{\mu,\nu}(\vec{k}).
\end{equation}

If we can make a gauge choice where $\chi_\mu(\vec{k})=\chi_\nu(\vec{k})$ for all relevant bands $\mu,\nu$, then $\vec{\mathcal{A}}^{-,+}(\vec{k})$ is an anti-symmetric function, as required for TR invariance. Note however that the value of $\chi_\mu(\vec{k}_{\rm TR})=0,\pi$ at TR invariant momenta $\vec{k}_{\rm TR} \equiv - \vec{k}_{\rm TR}  ~ {\rm mod}~ \vec{G}$, with $\vec{G}$ a reciprocal lattice vector, is fixed and can not be changed by gauge transformations. In this case, $e^{i \chi_\mu(\vec{k}_{\rm TR})} = \xi_\mu(\vec{k}_{\rm TR}) = \pm 1$ denotes the parity eigenvalues at $\vec{k}_{\rm TR}$. Therefore, in addition to inversion symmetry, we require that
\begin{equation}
\xi_\mu(\vec{k}_{\rm TR})=\xi_\nu(\vec{k}_{\rm TR}) \quad \text{for all} ~  \mu, \nu,
\label{eq:xiMuNuCond}
\end{equation}
at TR invariant momenta. As we later show, this condition can be dropped for particles in a spin-dependent artificial magnetic field.

The effective Hamiltonian in the $\ket{ \! \! \Uparrow}$-$\ket{\! \! \Downarrow}$ subspace is derived in detail in the supplementary material (SM). When only MW beam $j$ (with phase $\phi_{\rm MW}^{(j)}$) is switched on, we obtain
\begin{multline}
\H^{(j)} = \epsilon(\vec{k}) + \frac{|R_1(\vec{k})|^2 \ket{\! \! \Downarrow} \bra{ \Downarrow \! \!} + |R_2(\vec{k})|^2 \ket{\! \! \Uparrow} \bra{ \Uparrow \! \!}}{\Delta_{\vec{k}}} \\
+\frac{(\hbar \Omega_{\rm MW}^{(j)})^2}{4 \Delta_{\vec{k}}} \l \delta_{j,1} \ket{\! \! \Downarrow} \bra{ \Downarrow \! \!}  + \delta_{j,2} \ket{\! \! \Uparrow} \bra{\Uparrow \! \!} \r \\
+ \frac{\hbar \Omega_{\rm MW}^{(j)}}{2 \Delta_{\vec{k}}} \left[ \ket{ \! \! \Downarrow } \bra{\Uparrow \! \! } e^{- i \phi_{\rm MW}^{(j)}} R_j^*(\vec{k})  + \hc \right],
\label{eq:IdealizedSOC}
\end{multline}
where $R_{j} =  \frac{1}{2} \bigl[ F_{x} \mathcal{A}_x^{+,-} + F_{y} e^{i \phi_F^{(j)}} \mathcal{A}_y^{+,-} \bigr]$ with $\phi_F^{(1)}=\phi_F=-\phi_F^{(2)}$. The first two lines in Eq.\eqref{eq:IdealizedSOC} describe the free dispersion, which is renormalized by spin-dependent AC Stark shifts. If $\vec{\mathcal{A}}^{+,-}(\vec{k})$ is anti-symmetric, the last line in Eq.\eqref{eq:IdealizedSOC} describes the desired TR invariant SOC.

We proceed by considering the specific Berry connection for a spin-dependent square lattice,
\begin{equation}
\vec{\mathcal{A}}^{+,-}(\vec{k}) = \l A_x \sin k_x, A_y \sin k_y \r^T, \quad A_{x,y} \in \mathbb{R}.
\label{eq:Asquare}
\end{equation}
Note that $|R_1(\vec{k})|^2=|R_2(\vec{k})|^2=|R(\vec{k})|^2$ in this case. The last line of Eq.\eqref{eq:IdealizedSOC} now becomes
\begin{equation}
\H_{\rm SOC}^{(j)}(\vec{k}) = \frac{\hbar \Omega_{\rm MW}^{(j)}}{4 \Delta_{\vec{k}}} (\sin k_x, \sin k_y)  \l
\begin{array}{cc}
  g^{(j)}_x & -f^{(j)}_y \\
  f^{(j)}_x & - g^{(j)}_y
\end{array} \r  \l
\begin{array}{c}
  \hat{\sigma}^x \\
   \hat{\sigma}^y
\end{array} \r,
\end{equation}
with tunable amplitudes given by
\begin{flalign*}
&g^{(j)}_x = F_x A_x \cos ( \phi_{\rm MW}^{(j)} ), & f^{(j)}_x = F_y A_y \cos (\phi_{\rm MW}^{(j)} + \phi_F^{(j)}),  \\
&f^{(j)}_y =  F_x A_x \sin ( \phi_{\rm MW}^{(j)}), & g^{(j)}_y = F_y A_y \sin (\phi_{\rm MW}^{(j)} + \phi_F^{(j)}).
\end{flalign*}

With only one microwave drive, the spin-dependent AC Stark shifts in Eq.\eqref{eq:IdealizedSOC} break TR invariance. The simplest way to restore TR symmetry is to switch on both MW beams with equal Rabi frequencies $\Omega_{\rm MW}^{(1)}=\Omega_{\rm MW}^{(2)}$. In this case, however,  we can realize only equal-strength Rashba and Dresselhaus SOC due to interference effects between the two paths $j=1$ and $2$ (see SM). 

Alternatively, TR symmetry can be restored for an effective Floquet Hamiltonian when the two MW beams are switched on and off in an alternating fashion with frequency $\omega_s$. By choosing equal MW parameters ($\phi_{\rm MW}^{(1)}= \phi_{\rm MW}^{(2)}$, $\Omega_{\rm MW}^{(1)}= \Omega_{\rm MW}^{(2)}$) and switching the phase of the lattice shaking $\phi_F$ between $\phi_F^0$ while $\Omega_{\rm MW}^{(1)} \neq 0$ is on, and $-\phi_F^0$ while $\Omega_{\rm MW}^{(2)} \neq 0$ is on, the  spin-orbit part of the Hamiltonian becomes time-independent. For the choice $\phi_{\rm MW}=-\phi_F^0 = \pi/2$, this realizes tunable Rashba- and Dresselhaus couplings  
\begin{equation}
\alpha_{\rm R,D} =  \frac{\hbar \Omega_{\rm MW}}{8 \Delta} \l F_x A_x \pm F_y A_y \r,
\label{eq:alphaRDideal}
\end{equation}
where we approximated $\Delta_{\vec{k}} \approx \Delta$.

To lowest order in the Magnus expansion in $1/\omega_s$, the effective Floquet Hamiltonian contains SOC with amplitudes $\alpha_{\rm R,D}$ given in Eq. \eqref{eq:alphaRDideal}, and the AC Stark shift becomes $(|R(\vec{k})|^2 + |\hbar \Omega_{\rm MW}|^2/8 ) / \Delta_{\vec{k}}$, independent of the spin. More generally, we show in the SM that the effective Floquet Hamiltonian is TR invariant to all orders in $1/\omega_s$.

\begin{figure}[b!]
\centering
\epsfig{file=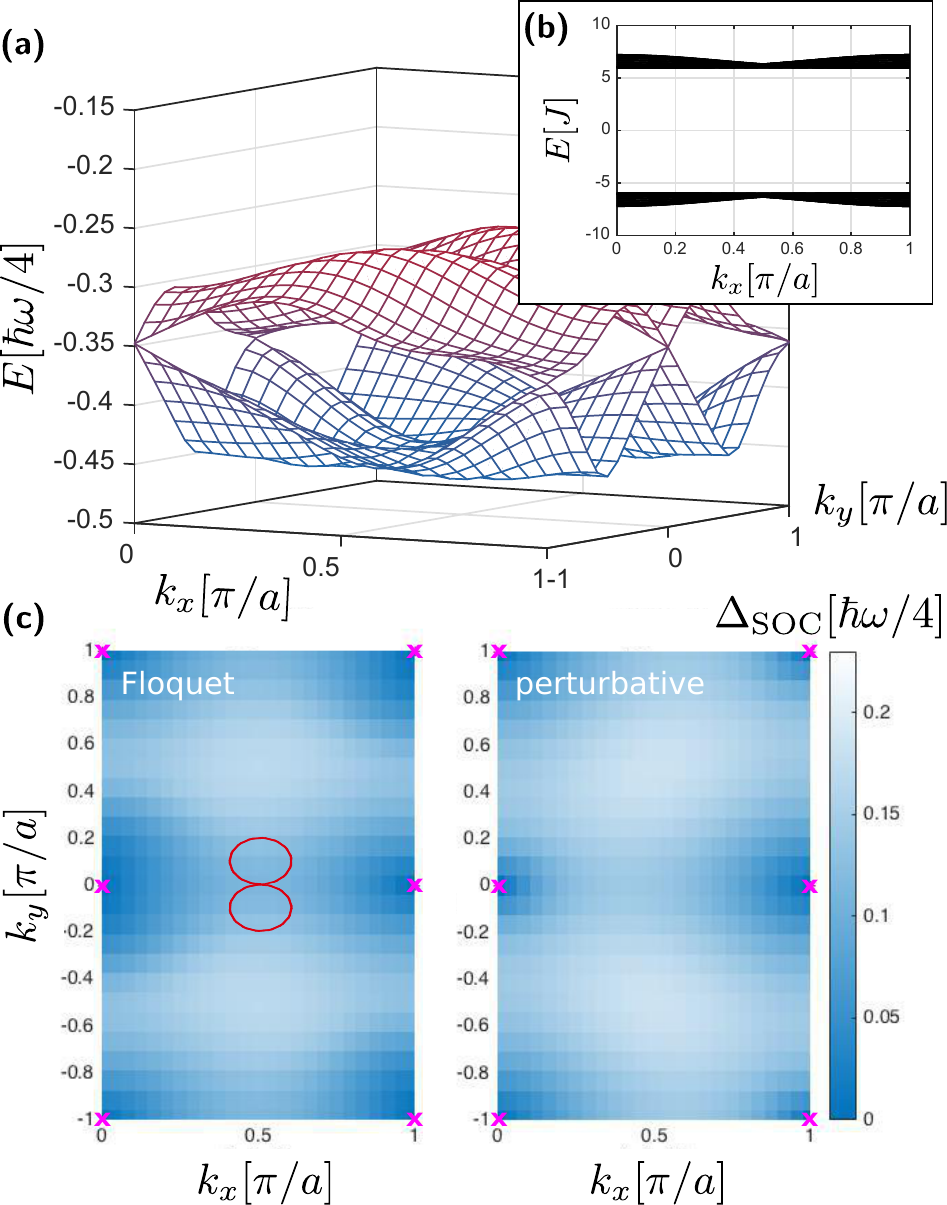, width=0.4\textwidth}
\caption{SOC in a spin-dependent square lattice: (a) We show the full band structure of the Floquet Hamiltonian (in units of $\hbar \omega_s = \hbar \omega/4 = 4 J$). Kramer's degeneracies at TR invariant momenta (crosses in (c)) are a direct consequence of the TR invariance of the system. (b) The bands of the free Hamiltonian are shown for $\delta=6 J$, as chosen in the numerics. In (c), we compare the value of the gap between the lowest two bands $\Delta_{\rm SOC}$ (due to SOC) obtained from the exact Floquet calculation and from the perturbative result in Eq.\eqref{eq:IdealizedSOC}. The momentum dependence $\vec{k}(t)$ due to lattice shaking is indicated by red circles. We set $\hbar \Omega_{\rm MW} = 5 J$ and $\omega_0 = 100 \omega$.}
\label{SquareLatticeSOC}
\end{figure}

\emph{Realistic scheme.--}
A realistic implementation of our scheme is possible using a square optical lattice with hopping amplitude $J$, superimposed by a spin-dependent superlattice with offset $\pm \delta \hat{\sigma}^z$ on the respective sublattices (see Fig.~\ref{IntroSOC} (b)). The resulting Bloch Hamiltonian is particle-hole symmetric and reads $\H_0(\vec{k}) = \epsilon(\vec{k}) ~ \hat{\tau}^z \otimes \hat{\sigma}^z$ in its eigenbasis $\ket{\pm_z}$. In the limit of a deep superlattice ($\delta \gg J$) this realizes a reversed band structure, and we obtain $\epsilon(\vec{k}) = - (\delta + \beta_{\vec{k}}^2 4 J^2/ \delta)$, where $\beta_{\vec{k}} = \cos(k_x)+\cos(k_y)$, and the Berry connection becomes spin-dependent, i.e, $\vec{\mathcal{A}}^{+,-}(\vec{k}) \propto \hat{\sigma}^z$ (see SM). In the limit $\delta \gg J$ we obtain $\vec{\mathcal{A}}^{+,-}(\vec{k}) = \hat{\sigma}^z (\sin k_x, \sin k_y)^T J / \delta$, as anticipated in Eq.\eqref{eq:Asquare}.

Tunable Rashba- and Dresselhaus SOC can be implemented as previously described using the model shown in Fig.~\ref{IntroSOC} (a). To account for the additional "$-$" sign due to the $\hat{\sigma}^z$-term in $\vec{\mathcal{A}}^{+,-}$, the phase $\phi_{\rm MW}^{(1)} = \phi_{\rm MW}^{(2)} + \pi$ needs to be shifted. The resulting SOC amplitudes are $\alpha_{\rm R,D} =  \lambda_{\rm SOC} \l F_x \pm F_y \r$, where $\lambda_{\rm SOC} = \hbar \Omega_{\rm MW} J/ ( 8 \delta \Delta)$.

In Fig.~\ref{SquareLatticeSOC} we present an exact calculation of the effective Floquet Hamiltonian in the square lattice for realistic experimental parameters. The band structure in Fig.~\ref{SquareLatticeSOC} (a) shows Kramer's degeneracies as a consequence of TR symmetry. The exact result closely resembles the perturbative calculation (based on Eq.\eqref{eq:IdealizedSOC}) shown on the same scale in Fig.~\ref{IntroSOC} (c). This agreement is confirmed by the comparison of the gap $\Delta_{\rm SOC}$ between the two lowest-lying bands in Fig.~\ref{SquareLatticeSOC} (c).

\begin{figure}[b!]
\centering
\epsfig{file=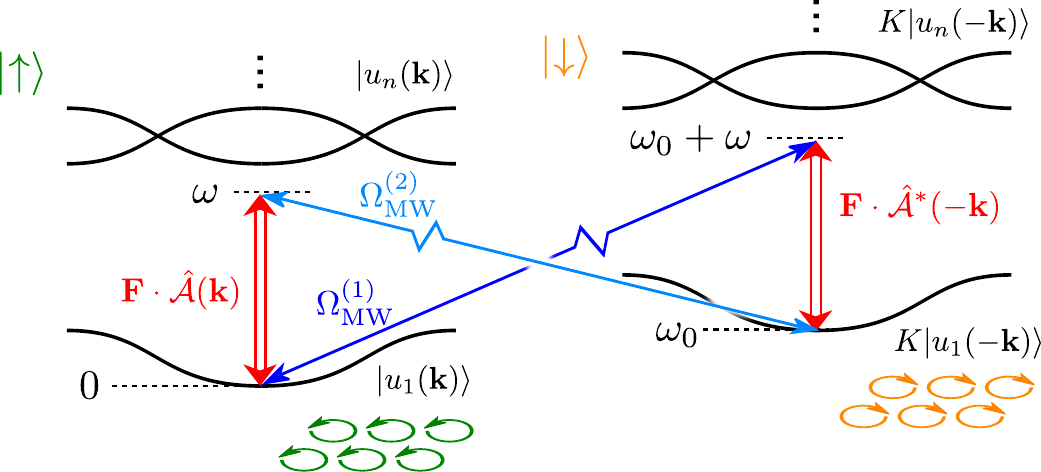, width=0.47\textwidth}
\caption{SOC in magnetic fields: Two identical but time-reversed band structures, with opposite Chern numbers \cite{Aidelsburger2013} for  spins $\ket{\!\! \uparrow}$ and $\ket{ \!\! \downarrow}$ can be coupled by microwave beams and lattice shaking. A two-photon process  through a virtually excited state allows for the realization of TR invariant SOC between the time-reversed bands.}
\label{MagneticFieldSOC}
\end{figure}

\emph{SOC in the presence of magnetic fields.--}
Now we demonstrate how SOC can be generated in more general situations. Our starting point are two identical but time-reversed copies of a Chern insulator ($\ket{ \! \uparrow}$ and $\ket{ \! \downarrow}$) with opposite Chern numbers $\mathcal{C}=\pm 1$. This situation was considered by Kane and Mele \cite{Kane2005}, who showed that even when SOC mixes the two systems, a $\mathbb{Z}_2$ topological invariant characterizing the topological insulator remains quantized as long as TR symmetry is retained. While two time-reversed copies of a system with equal but opposite Chern numbers have been realized experimentally using ultracold atoms \cite{Aidelsburger2013}, the effect of SOC has not yet been investigated in this context.

\begin{figure}[t!]
\centering
\epsfig{file=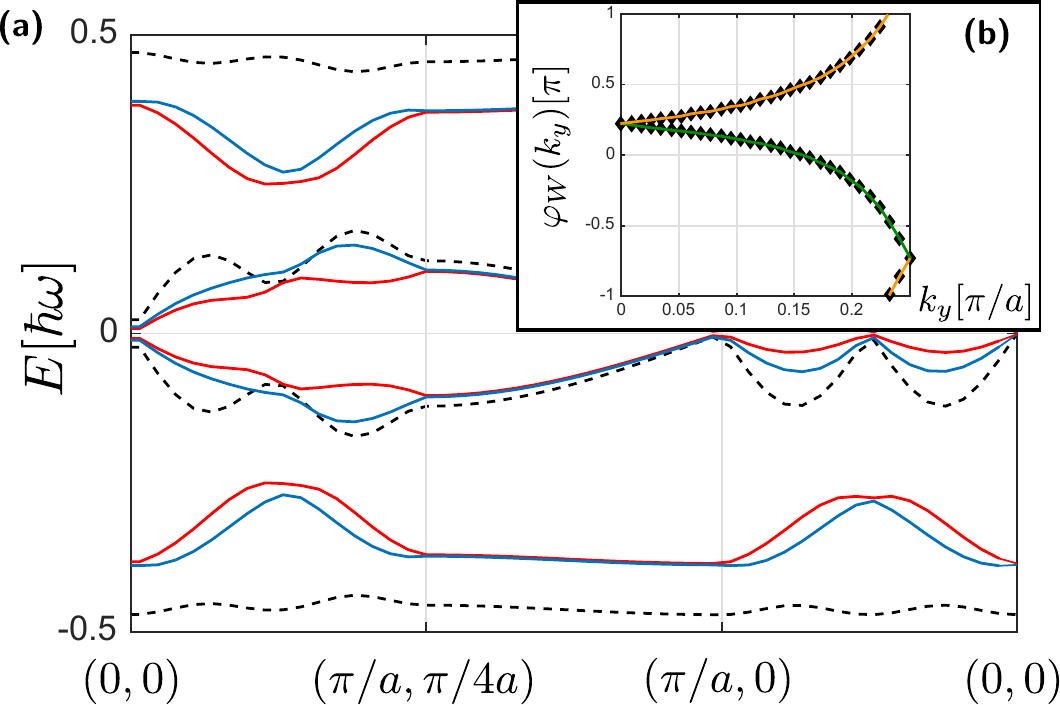, width=0.47\textwidth}
\caption{SOC in the Hofstadter model: (a) We start from two spin-degenerate time-reversed bands in the Hofstadter model (dashed lines) at magnetic flux per plaquette $\alpha = \pm 1/4$. Coupling the bands by MW beams and lattice shaking as described in the text leads to TR invariant SOC, which partly lifts the spin degeneracy (solid lines). The two lowest bands are a $\mathbb{Z}_2$ topological insulator, which can be seen from the eigenvalues $\exp (i \varphi_W)$ of the $U(2)$ Wilson loop (b). Parameters are $\hbar \omega=6 J$, $\hbar \omega_0=100 J$, $\hbar \Omega_{\rm MW}=5 J$, $\phi_{\rm MW}=\pi$ and $F_x=F_y=5 J/a$, where $J$ is the tunneling amplitude.}
\label{MagneticFieldSOCmain}
\end{figure}

A direct generalization of our scheme is shown in Fig.~\ref{MagneticFieldSOC}. As before, TR invariant SOC between the Bloch band $\ket{\! \! \Uparrow} = \ket{u_1(\vec{k}),\uparrow}$ and its TR partner $\ket{\! \! \Downarrow} = K \ket{u_1(-\vec{k}),\downarrow}$ at quasimomentum $\vec{k}$ is generated by a combination of MW transitions and lattice shaking. In contrast to the case of the reversed Bloch bands in the simplified model of Fig.~\ref{IntroSOC} (a), the MW transitions are renormalized by a gauge-dependent Franck-Condon overlap $K^{m,n}(\vec{k}) = \langle K u_m(-\vec{k}) \ket{u_n(\vec{k})}$. This allows us to drop condition \eqref{eq:xiMuNuCond} and assume only that the system has inversion symmetry (see Eq.\eqref{eq:relationInversion}). As a result (see SM for details) the effective Rabi coupling between $\ket{\! \! \Uparrow}$ and $\ket{\! \! \Downarrow}$ is an anti-symmetric function of $\vec{k}$, independent of the gauge choice. 

Here we only consider the case when both MW beams $\Omega_{\rm MW}^{(1,2)} = \Omega_{\rm MW}$ are switched on simultaneously. To avoid spin-dependent AC Stark shifts, we make the choice $\phi_F=0$, corresponding to linear lattice shaking. These conditions guarantee TR invariance of the resulting Hamiltonian (see SM and Ref.~\cite{Kitagawa2010}). Furthermore, we choose $\phi_{\rm MW}^{(1)} = \phi_{\rm MW}^{(2)}+\pi$ to obtain constructive interference between the two pathways.

In Fig.~\ref{MagneticFieldSOCmain}, we present an exact calculation of the band structure in a Hofstadter model \cite{HOFSTADTER1976,Aidelsburger2013}. When MW coupling and lattice shaking are switched on, the initial spin degeneracy is lifted by the presence of synthetic SOC. At TR invariant momenta, we obtain Kramer's degeneracies as a consequence of the symmetry. The resulting band structure is a $\mathbb{Z}_2$ topological insulator, which can be checked by calculating the winding of the $U(2)$ Wilson loops \cite{Yu2011}~ (see Fig.~\ref{MagneticFieldSOCmain}). Wilson loops can be directly measured \cite{Grusdt2014Z2,Li2016} to experimentally test our prediction.

\emph{Summary and Outlook.--}
In this letter we introduced a general scheme for realizing TR invariant synthetic 2D SOC in optical lattices. We made use of a combination of direct MW transitions and near-resonant lattice shaking, which provides the required momentum dependence in the effective Hamiltonian. Spin-dependent optical lattices provide a realistic platform where our scheme can be used to implement Rashba and Dresselhaus SOC with fully tunable strengths. We expect that this will enable an experimental investigation of exotic physics related to SOC in the near future, ranging from studies of supersymmetric Hamiltonians \cite{Tomka2015} to statistical transmutations induced by strong interactions \cite{Sedrakyan2012}.

Our scheme can also be used to introduce SOC in the presence of magnetic fields. In particular, it allows for the realization of TR invariant $\mathbb{Z}_2$ topological insulators with non-trivial spin textures, as in the celebrated Kane-Mele model \cite{Kane2005}. This will enable experiments with ultracold atom to explore topologically protected edge states, in a situation closely resembling realistic solids with strong interactions in the presence of SOC.

\emph{Acknowledgements.--}
The authors would like to thank M. Aidelsburger, A. Bohrdt, and M. Endres for fruitful discussions. The authors acknowledge support by the Gordon and Betty Moore foundation, Harvard-MIT CUA, NSF Grant No. DMR-1308435, AFOSR Quantum Simulation MURI, AFOSR MURI Photonic Quantum Matter and the Humboldt Foundation.

\appendix

\section{SOC Hamiltonian for the model with reversed Bloch bands}
\label{appdx:IdealCase}
In this supplementary we present a detailed derivation of the effective SOC Hamiltonian explained in the main text. We consider the idealized situation shown in Fig.~\ref{IntroSOC} (a), where the Bloch bands corresponding to spin up ($\ket{\! \uparrow}$) particles are obtained by reversing the order of the Bloch bands associated with the spin down ($\ket{\! \downarrow}$) degrees of freedom and shifting them in energy by an amount $\hbar \omega_0$. Without the additional couplings, the free Hamiltonian reads (we use $\hbar=1$ in this supplementary)
\begin{equation}
\H_0(\vec{k}) = \epsilon(\vec{k}) \hat{\sigma}^z \otimes \hat{\tau}^z + \frac{\omega_0}{2} \l 1 - \hat{\sigma}^z  \r.
\end{equation}
Here $\hat{\sigma}^z = \ket{ \! \uparrow} \bra{ \uparrow \!} - \ket{ \! \downarrow} \bra{ \downarrow \!}$ acts on (pseudo) spin degrees of freedom, and $\hat{\tau}^z = \ket{+} \bra{ + } - \ket{-} \bra{-}$ labels motional degrees of freedom associated with the band structure; The corresponding cell-periodic Bloch wavefunctions are $\ket{u_\pm(\vec{k})}$. 

We assume that pseudospins are offset in energy by $\omega_0$. To eliminate $\omega_0$ from the Hamiltonian, we introduce a rotating basis $\ket{\tilde{\downarrow}} = e^{- i \omega_0 t} \ket{ \! \downarrow}$. In the new frame the free part becomes
\begin{equation}
\H_0(\vec{k}) = \epsilon(\vec{k}) \hat{\sigma}^z \otimes \hat{\tau}^z,
\end{equation}
and to keep the notation compact we replace $\ket{ \tilde{\downarrow}}$ by $\ket{ \! \downarrow}$.

\emph{Lattice shaking.--} As a first step towards SOC, we introduce couplings between the two bands $\ket{u_\pm(\vec{k})}$ by lattice shaking. To this end an oscillating force is applied, which will be parametrized by
\begin{equation}
\vec{F}(t) = \vec{e}_x F_{x} \cos(\omega t) + \vec{e}_y F_{y} \cos(\omega t + \phi_F(t)).
\label{eq:FtParam}
\end{equation}
We assume that the driving frequency $\omega$ is close to, but not in resonance with, the band gap $2 |\epsilon(\vec{k})|$; $\phi_F(t)$ denotes the phase of the driving. As a result of this force, the quasimomentum of particles in the Brillouin zone changes in time,
\begin{equation}
\vec{k}(t) = \vec{k}(0) - \int_0^t d\tau ~ \vec{F}(\tau).
\end{equation}

For simplicity we assume that the Bloch oscillation frequency is small compared to the shaking frequency, 
\begin{equation}
|\vec{F}| a \ll \omega,
\end{equation}
allowing to approximate $\vec{k}(t) \approx \vec{k}(0) \equiv \vec{k}$. Then we obtain the following Hamiltonian describing the effect of the oscillatory force, see e.g. Refs. \cite{Grusdt2014Z2,Dunlap1986,Li2016},
\begin{equation}
\H_{F}(\vec{k},t) = \vec{F}(t) \cdot \vec{\mathcal{\hat{A}}}(\vec{k}).
\end{equation}

Next we make use of the rotating wave approximation (RWA), which is justified because $|\vec{\mathcal{\hat{A}}} \cdot \vec{F}| \simeq |\vec{F}| a \ll \omega$ and since we assume that $\omega \approx 2 |\epsilon(\vec{k})|$ is comparable to the band gap. We use the decomposition 
\begin{equation}
\vec{\mathcal{\hat{A}}}(\vec{k}) = \vec{\mathcal{A}}^x(\vec{k}) \hat{\tau}^x+\vec{\mathcal{A}}^y(\vec{k}) \hat{\tau}^y+\vec{\mathcal{A}}^z(\vec{k}) \hat{\tau}^z,
\end{equation}
where $\vec{\mathcal{A}}^\mu$ are real vectors ($\mu=x,y,z$) with components $\mathcal{A}_i^\mu$ ($i=x,y$). For spin $\ket{ \! \uparrow}$, where $\ket{+}$ is lower in energy than $\ket{-}$, the RWA result is
\begin{equation}
\vec{F}(t) \cdot \vec{\mathcal{\hat{A}}}(\vec{k}) \stackrel{\rm RWA}{\approx} R(\vec{k},\phi_F(t)) e^{i \omega t} \ket{+,\uparrow} \bra{-,\uparrow} + \hc,
\end{equation}
where $R(\vec{k},\phi_F)$ is defined by
\begin{equation}
R(\vec{k},\phi_F) =  \frac{1}{2} \left[ F_{x} \l \mathcal{A}_x^x - i \mathcal{A}_x^y \r + F_{y} e^{i \phi_F} \l  \mathcal{A}_y^x -i \mathcal{A}_y^y \r \right].
\end{equation}
For spin $\ket{ \! \downarrow}$ on the other hand, where $\ket{+}$ is higher in energy than $\ket{-}$, the RWA result is
\begin{equation}
\vec{F}(t) \cdot \vec{\mathcal{\hat{A}}}(\vec{k}) \stackrel{\rm RWA}{\approx} R(\vec{k},-\phi_F(t)) e^{- i \omega t} \ket{+,\downarrow} \bra{-,\downarrow} + \hc
\end{equation}
Note that within the RWA we may also neglect the terms $\vec{F}(t) \cdot \vec{\mathcal{A}}^z(\vec{k})$ which are oscillating at frequency $\omega$.

\emph{MW couplings.--} To obtain an effective SOC Hamiltonian for the two pseudospin states 
\begin{equation}
\ket{\! \Uparrow, \vec{k}} = \ket{\! \uparrow} \ket{u_+(\vec{k})}, \qquad  \ket{\! \Downarrow, \vec{k}} = \ket{\! \downarrow} \ket{u_-(\vec{k})},
\end{equation}
we include MW couplings between different spins. They will only be switched on one at a time to avoid interference effects, and their frequencies are chosen such that $\ket{ \! \! \Uparrow}$ and $\ket{\! \! \Downarrow}$ are resonantly coupled by a second order process. The Hamiltonians for these two processes ($j=1,2$) are given by
\begin{equation}
\H_{\rm MW}^{(j)}(t) = \Omega_{\rm MW}^{(j)} \cos \l \omega_j t + \phi_{\rm MW}^{(j)} \r  \hat{\sigma}^x,
\end{equation}
where $\omega_1 = \omega_0 + \omega$ and $\omega_2 = \omega_0 - \omega$. We assume that
\begin{equation}
\omega_0 \gg \omega
\label{eq:omega0ggomega}
\end{equation}
such that $\omega_{1,2} > 0$ before going to the frame rotating with frequency $\omega_0$.

In the frame rotating with frequency $\omega_0$, on the other hand, $\omega_{1,2}$ are replaced by $\omega_{1,2}=\pm \omega$. The last condition Eq.~\eqref{eq:omega0ggomega} together with the assumption
\begin{equation}
|\Omega_{\rm MW}^{(j)}| \ll \omega, \qquad j=1,2
\end{equation}
justifies using the RWA, from which we obtain
\begin{flalign}
\H_{\rm MW}^{(1)}(t)  &\stackrel{\rm RWA}{\approx}  \frac{\Omega_{\rm MW}^{(1)}}{2} e^{i ( \omega_1 t + \phi_{\rm MW}^{(1)} ) }  \ket{+, \uparrow} \bra{+, \downarrow \!} + \hc ~,\\
\H_{\rm MW}^{(2)}(t)  &\stackrel{\rm RWA}{\approx}  \frac{\Omega_{\rm MW}^{(2)}}{2} e^{i ( \omega_2 t + \phi_{\rm MW}^{(2)} ) }  \ket{ -, \uparrow} \bra{-, \downarrow \!} + \hc ~ .
\end{flalign}

\emph{Hamiltonian in RWA.--}
Combining the terms derived above by applying the RWA leads to the following Hamiltonian, formulated in the frame rotating at frequency $\omega_0$. If the MW beam $j=1$ is switched on we obtain
\begin{multline}
\H_{\rm tot}^{(1)}(\vec{k})  = \H_0(\vec{k}) + \biggl[ \frac{\Omega_{\rm MW}^{(1)}}{2} e^{i ( \omega_1 t + \phi_{\rm MW}^{(1)} ) }  \ket{+, \uparrow} \bra{+, \downarrow \!} \\
+R(\vec{k},\phi_F(t)) e^{i \omega t} \ket{+,\uparrow} \bra{-,\uparrow}\\
+ R(\vec{k},-\phi_F(t)) e^{- i \omega t} \ket{+,\downarrow} \bra{-,\downarrow} + \hc \biggr],
\end{multline}
and a similar expression is derived when MW beam $j=2$ is switched on.

\emph{Effective Hamiltonian.--}
When both lattice shaking and one of the MW drivings ($j$) is switched on, we obtain an effective Hamiltonian coupling $\ket{ \! \! \Uparrow}$ and $\ket{\! \! \Downarrow}$ from second order perturbation theory and within RWA. In the frame rotating with frequency $\omega_0$ it reads
\begin{multline}
\H^{(j)} = \epsilon(\vec{k}) + \frac{|R_1(\vec{k})|^2 \ket{\! \! \Downarrow} \bra{ \Downarrow \! \!} + |R_2(\vec{k})|^2 \ket{\! \! \Uparrow} \bra{ \Uparrow \! \!}}{\Delta_{\vec{k}}} \\
+\frac{(\Omega_{\rm MW}^{(j)})^2}{4 \Delta_{\vec{k}}} \l \delta_{j,1} \ket{\! \! \Downarrow} \bra{ \Downarrow \! \!}  + \delta_{j,2} \ket{\! \! \Uparrow} \bra{\Uparrow \! \!} \r \\
+ \frac{\Omega_{\rm MW}^{(j)}}{2 \Delta_{\vec{k}}} \left[ \ket{ \! \! \Downarrow } \bra{\Uparrow \! \! } e^{- i \phi_{\rm MW}^{(j)}} R_j^*(\vec{k})  + \hc \right],
\label{eq:EffHamiltonianIdeal}
\end{multline}
Here $\Delta_{\vec{k}} = 2 \epsilon(\vec{k}) + \omega$ and we introduced $R_1(\vec{k}) = R(\vec{k},\phi_F(t))$ and $R_2(\vec{k}) = R(\vec{k},-\phi_F(t))$. The first line includes the bare dispersion $\epsilon(\vec{k})$, AC Stark shifts for both pseudospins from lattice shaking as well as MW driving. 

\emph{Floquet Hamiltonian.--}
The spin-dependent AC Stark shift corresponds to a broken TR symmetry. In order to eliminate it, we consider a pulsed sequence where the two MW beams $j=1,2$ are switched on and off in an alternating fashion. In the first interval, from $t=0$ to $T_s/2$, only the $j=1$ beam is operating, and from $t=T_s/2$ to $t=T_s$ only the $j=2$ beam is switched on. When $j=2$ is acting on the system, the phase of the driving force is reversed, i.e. $\phi_F(t)=-\phi_F^0$ for $T_s/2 < t < T_s$ and $\phi_F(t) = \phi_F^0$ for $0<t < T_s/2$. The switching frequency is defined by $\omega_s = 2 \pi / T_s$. These dynamics give rise to an effective Floquet Hamiltonian $\H_{\rm eff}$ which we calculate in the limit of large $\omega_s$ using the Magnus expansion. To lowest order we obtain for $\Omega_{\rm MW}^{(1)} = \Omega_{\rm MW}^{(2)} = \Omega_{\rm MW}$ and $\phi_{\rm MW}^{(1)}=\phi_{\rm MW}^{(2)}=\phi_{\rm MW}$
\begin{multline}
\H^{(0)}_{\rm eff}(\vec{k}) = \epsilon(\vec{k}) + \frac{|R_1(\vec{k})|^2 + |R_2(\vec{k})|^2 + (\Omega_{\rm MW})^2/4}{2 \Delta_{\vec{k}}}  + \\
+ \frac{\Omega_{\rm MW}}{2 \Delta_{\vec{k}}} \left[ \ket{ \! \! \Downarrow } \bra{\Uparrow \! \! } e^{- i \phi_{\rm MW}} R^*(\vec{k},\phi_F^0)  + \hc \right],
\end{multline}
where the first line describes how the dispersion relation is renormalized. The second line represents SOC and can be written as
\begin{multline}
\H_{\rm SOC}(\vec{k}) =  \frac{\Omega_{\rm MW}}{2 \Delta_{\vec{k}}} \Bigl( \hat{\sigma}^x ~ {\rm Re} \left[ e^{i \phi_{\rm MW}} R(\vec{k},\phi_F^0) \right] \\
- \hat{\sigma}^y ~ {\rm Im} \left[ e^{i \phi_{\rm MW}} R(\vec{k},\phi_F^0) \right] \Bigr).
\end{multline}
This term is TR invariant in the presence of the following symmetry
\begin{equation}
R(-\vec{k},\phi_F^0) = - R(\vec{k},\phi_F^0),
\end{equation}
as discussed in the main text. The properties of the resulting SOC depend on off-diagonal elements of the Berry connection $\vec{\mathcal{\hat{A}}}(\vec{k})$ and can be tuned by the phases $\phi_F^0$ of the shaking and $\phi_{\rm MW}$ of the MW beams.

Higher order terms in the Magnus expansion give rise to corrections in the effective Floquet Hamiltonian, that may be relevant for the effective band structure. Now we proof that such corrections are TR invariant to all orders if the SOC Hamiltonian has this property. The effective Hamiltonian $\H_{\rm eff}^{(\infty)}$ is defined as
\begin{equation}
\hat{U}_{\vec{k}} = e^{ - i T_s \H_{\rm eff}^{(\infty)}(\vec{k})} = e^{-i \H^{(2)}(\vec{k}) T_s/2 } e^{-i \H^{(1)}(\vec{k}) T_s/2 },
\end{equation}
and it is TR invariant if $\hat{\theta}^\dagger \H_{\rm eff}^{(\infty)}(\vec{k}) \hat{\theta} = \H_{\rm eff}^{(\infty)}(-\vec{k})$, i.e. for $\hat{\theta}^\dagger \hat{U}_{\vec{k}} \hat{\theta} = \hat{U}^\dagger_{- \vec{k}}$. Because by construction $\hat{\theta}^\dagger \H^{(2)}(\vec{k}) \hat{\theta} = \H^{(1)}(-\vec{k})$ it follows that
\begin{equation}
\hat{\theta}^\dagger \hat{U}_{\vec{k}} \hat{\theta} = e^{i \H^{(1)}(-\vec{k}) T_s/2 } e^{i \H^{(2)}(-\vec{k}) T_s/2 } = \hat{U}_{-\vec{k}}^\dagger.
\end{equation}

\section{SOC in a spin-dependent square lattice}
In this SM we apply our idealized scheme to the realistic situation of a spin-dependent square lattice. As described in the main text, the Hamiltonian consists of nearest-neighbor hopping with amplitude $J$, and a spin-dependent superlattice potential generating energy offsets $\pm \delta$ for opposite spins and different sublattices. In second quantization, the Hamiltonian reads
\begin{equation}
\H = - J \sum_{\langle i,j \rangle} \l  \ad_{i,\sigma} \a_{j,\sigma} + \hc \r + \delta \sum_j (-1)^\sigma (-1)^j \ad_{j,\sigma} \a_{j,\sigma},
\end{equation}
where $(-1)^j=-1 (+1)$ for $j$ from the $A$ ($B$) sublattice, and similarly $(-1)^\downarrow=-1$ and $(-1)^\uparrow=1$. 

\emph{Band structure.--}
The Bloch Hamiltonian can conveniently be written
\begin{equation}
\H_0(\vec{k}) = \delta ~ \hat{\sigma}^z \otimes \tilde{\tau}^z - 2 J \beta_{\vec{k}} \tilde{\tau}^x,
\end{equation}
where $\beta_{\vec{k}} = \cos(k_x)+\cos(k_y)$ and the Pauli matrix $\tilde{\tau}^z$ is defined in the basis $\ket{A}$, $\ket{B}$ of the $A$ and $B$ sublattice. Because $\{ \H_0(\vec{k}) , \tilde{\tau}^y \} = 0$, the Hamiltonian is particle-hole symmetric and we can write $\H_0(\vec{k}) = \epsilon(\vec{k},\sigma) ~  \hat{\tau}^z$, where $\hat{\tau}^z$ is defined in the eigenbasis. 

The cell-periodic Bloch states $\ket{u_{\vec{k},\sigma}^\pm}$, with $\hat{\tau}^z \ket{u_{\vec{k},\sigma}^\pm} = \pm \ket{u_{\vec{k},\sigma}^\pm}$, can conveniently be written in the eigenbasis $\ket{\pm_y}$ of $\hat{\tau}^y \equiv \tilde{\tau}^y$,
\begin{equation}
\ket{u_{\vec{k},\sigma}^\pm} = \l \ket{+_y} \pm e^{i \vartheta_{\vec{k},\sigma}} \ket{-_y} \r / \sqrt{2}.
\label{eq:ukSigmaPM}
\end{equation}
It holds $\epsilon(\vec{k},\sigma) = - \hat{\sigma}^z \delta \cos \vartheta_{\vec{k},\sigma} + 2 J \beta_{\vec{k}} \sin \vartheta_{\vec{k},\sigma}$, where the mixing angle is given by $\tan \vartheta_{\vec{k},\sigma} = - \hat{\sigma}^z 2 J \beta_{\vec{k}} / \delta$. Hence we can write the Hamiltonian as
\begin{equation}
\H_0(\vec{k}) =  \epsilon(\vec{k}) ~ \hat{\sigma}^z \otimes \hat{\tau}^z,
\end{equation}
where we defined
\begin{flalign}
\epsilon(\vec{k}) &= -  \delta \cos \vartheta_{\vec{k}} + 2 J \beta_{\vec{k}} \sin \vartheta_{\vec{k}},\\
\tan \vartheta_{\vec{k}} &= - 2 J \beta_{\vec{k}} / \delta.
\end{flalign}
This shows that the band energies are completely inverted for different spins.  

The Bloch wavefunctions are completely inverted, $\ket{u_{\vec{k},\uparrow}^+} = \ket{u_{\vec{k},\downarrow}^+}$, only for a deep superlattice $\delta \gg J$. In the other limit $J \gg \delta$ there is no inversion and it holds $\ket{u_{\vec{k},\uparrow}^+} = \ket{u_{\vec{k},\downarrow}^-}$. As a consequence, the $U(2)$ Berry connection becomes spin dependent. From Eq.\eqref{eq:ukSigmaPM} we obtain the exact result
\begin{equation}
\vec{\mathcal{A}}^{+,-}(\vec{k}) = - \hat{\sigma}^z \frac{J}{\delta} \cos^2 \vartheta_{\vec{k}} \vec{\nabla}_{\vec{k}} \beta_{\vec{k}}.
\label{eq:ApmGeneralRes}
\end{equation}
Using this gauge choice, it follows immediately that $\vec{\mathcal{A}}^{+,-}(\vec{k}) = - \vec{\mathcal{A}}^{+,-}(-\vec{k})$ is antisymmetric. Thus $\xi_+(\vec{k}_{\rm TR}) = \xi_-(\vec{k}_{\rm TR})$ as required for TR invariance of the effective SOC Hamiltonian. We also note that $\vec{\mathcal{A}}^{+,-}(\vec{k}) \propto \hat{\sigma}^z$ to all orders in $J/\delta$. Because $\cos \vartheta_{\vec{k}} = 1 + \mathcal{O}(J/\delta)^2$ we obtain
\begin{equation}
\vec{\mathcal{A}}^{+,-}(\vec{k}) = \hat{\sigma}^z \frac{J}{\delta} \l \sin k_x , \sin k_y \r^T + \mathcal{O}(J/\delta)^3.
\label{eq:ApmPert}
\end{equation}

\emph{Effective SOC.--}
As described in the main text, our method for generating tunable SOC can be applied to the spin-dependent square lattice. Taking into account the additional spin-dependence of the Bloch states leads to the following effective Hamiltonian, 
\begin{multline}
\H^{(j)} = \epsilon(\vec{k}) + \frac{|R_1(\vec{k})|^2 \ket{\! \! \Downarrow} \bra{ \Downarrow \! \!} + |R_2(\vec{k})|^2 \ket{\! \! \Uparrow} \bra{ \Uparrow \! \!}}{\Delta_{\vec{k}}} \\
+\frac{(\Omega_{\rm MW}^{(j)})^2 |\lambda|^2}{4 \Delta_{\vec{k}}} \l \delta_{j,1} \ket{\! \! \Downarrow} \bra{ \Downarrow \! \!}  + \delta_{j,2} \ket{\! \! \Uparrow} \bra{\Uparrow \! \!} \r \\
+ (-1)^j ~ \frac{\Omega_{\rm MW}^{(j)}}{2 \Delta_{\vec{k}}} \left[ \ket{ \! \! \Downarrow } \bra{\Uparrow \! \! } e^{- i \phi_{\rm MW}^{(j)}} \lambda R_j^*(\vec{k})  + \hc \right].
\label{eq:EffHamiltonianSquareLattice}
\end{multline}
The only difference to Eq.\eqref{eq:EffHamiltonianIdeal} is the additional minus sign in the last line, due to the factor $\hat{\sigma}^z$ in Eq.\eqref{eq:ApmGeneralRes}, and the appearance of Franck-Condon factors
\begin{equation}
\lambda = \langle u_{\vec{k},\downarrow}^\pm \ket{u_{\vec{k},\uparrow}^\pm} = \frac{1}{2} \l 1 + e^{2 i \vartheta_{\vec{k}}} \r.
\end{equation}
Note that we defined $R_j(\vec{k})$ using the expression for $\vec{\mathcal{A}}^{+,-}(\vec{k})$ with $\hat{\sigma}^z=1$.

Because $\lambda$ is independent of the path $j$, the methods discussed for the idealized situation still guarantee a TR symmetric effective Hamiltonian. To leading order in $J/\delta$ the Franck-Condon factors are equal to one, $\lambda = 1 - 2 i \beta_{\vec{k}} J / \delta + \mathcal{O}(J/\delta)^2$. Therefore in the limit $J \ll \delta$ the results from Appendix \ref{appdx:IdealCase} carry over directly, if $\phi_{\rm MW}^{(1)} \to \phi_{\rm MW}^{(1)} + \pi$ is shifted by $\pi$ to compensate for the additional minus sign $(-1)^j$ in the last line of Eq.\eqref{eq:EffHamiltonianSquareLattice}. $\lambda$ merely renormalizes the coupling strength of SOC.

\emph{Equal Rashba- and Dresselhaus SOC.--}
Here we discuss how equal Rashba and Dresselhaus SOC can be generated in the spin-dependent square optical lattice. To this end both MW beams are switched on with equal strengths, $\Omega_{\rm MW}^{(1,2)} = \Omega_{\rm MW}$. For simplicity we work in the limit $J \ll \delta$, where $\vec{\mathcal{A}}^{+,-}(\vec{k}) $ is real, see Eq.\eqref{eq:ApmPert}, AC-Stark shifts are independent of the spin, and $\lambda \approx 1$.

The effective Hamiltonian becomes,
\begin{equation}
\H_{\rm eff} = \epsilon(\vec{k}) + \frac{|R(\vec{k})|^2 + (\Omega_{\rm MW})^2/4}{\Delta_{\vec{k}}} + \H_{\rm SOC}(\vec{k}),
\label{eq:EffHamiltonianEqualRD}
\end{equation}
where the SOC is described by
\begin{flalign}
\H_{\rm SOC}(\vec{k}) &=  \frac{\Omega_{\rm MW}}{2 \Delta_{\vec{k}}} \left[ \hat{\sigma}^x ~ {\rm Re} (\eta_{\vec{k}}) - \hat{\sigma}^y ~ {\rm Im} (\eta_{\vec{k}}) \right], \\
\eta_{\vec{k}} &=e^{i \phi_{\rm MW}^{(2)}} R(\vec{k},-\phi_F) - e^{i \phi_{\rm MW}^{(1)}} R(\vec{k},\phi_F).
\end{flalign}
Due to interference effects between the two beams, only equal Rashba-Dresselhaus SOC can be generated in this way. For example, the choice $\phi_{\rm MW}^{(1)}=\phi_{\rm MW}^{(2)}=\phi_{\rm MW}$ leads to
\begin{multline}
\H_{\rm SOC}(\vec{k}) =  \frac{\Omega_{\rm MW}}{2 \Delta_{\vec{k}}} \frac{J}{\delta} F_y \sin (\phi_F) \\
\times  \sin (k_y) \left[ \cos (\phi_{\rm MW}) \hat{\sigma}^y + \sin (\phi_{\rm MW}) \hat{\sigma}^x \right].
\end{multline}

\section{SOC in the presence of magnetic fields}
In this supplementary we present a detailed derivation of the effective SOC Hamiltonian in the presence of (artificial) magnetic fields. We consider the situation shown in Fig.~\ref{MagneticFieldSOC}, where the band structure of spin down ($\ket{ \! \! \downarrow}$) particles is obtained from the up spins ($\ket{ \! \! \uparrow}$) by time-reversal (i.e. complex conjugation in this case). Without the additional couplings, the free Hamiltonian reads ($\hbar=1$)
\begin{multline}
\H_0(\vec{k}) = \ket{\! \uparrow} \bra{\uparrow \!} \sum_n \epsilon_n(\vec{k}) \ket{u_n(\vec{k})} \bra{u_n(\vec{k})} +\\
+ \ket{\! \downarrow} \bra{\downarrow \!} \sum_n \epsilon_n(\vec{k}) K \ket{u_n(-\vec{k})} \bra{K u_n(-\vec{k})},
\end{multline}
where $\ket{u_n(\vec{k})}$ denotes the magnetic Bloch wavefunctions of $\H_\uparrow(\vec{k})$ with energies $\epsilon_n(\vec{k})$. For example, we can choose $\H_\uparrow(\vec{k})$ to describe the Haldane \cite{Haldane1988} or the Hofstadter model \cite{HOFSTADTER1976}. The Hamiltonian for spin down states is obtained by TR,
\begin{equation}
\H_\downarrow(\vec{k}) = K \H_\uparrow(-\vec{k}) K,
\end{equation}
and the corresponding eigenfunctions at energy $\epsilon_n(\vec{k})$ are given by $K \ket{u_n(-\vec{k})}$. As in the construction by Kane and Mele \cite{Kane2005}, the Hamiltonian $\H_0$ is TR invariant,
\begin{equation}
\hat{\theta}^\dagger \H_0(\vec{k}) \hat{\theta} = \H_0(-\vec{k}),
\end{equation}
where $\hat{\theta}= K i \hat{\sigma}^y$.

\emph{Lattice shaking.--} Next we introduce lattice shaking by applying an oscillatory force $\vec{F}(t)$ parametrized as in Eq.\eqref{eq:FtParam}. It couples the lowest band $n=1$, which we are interested in, to higher bands $n$,
\begin{multline}
\H_{F}(\vec{k},t) = \vec{F}(t) \cdot \vec{\mathcal{\hat{A}}}(\vec{k})  \stackrel{\rm RWA}{\approx}\\
 \stackrel{\rm RWA}{\approx} \sum_{n>1} \left[ R_n(\vec{k}) e^{i \omega t} \ket{\! \uparrow} \bra{\uparrow \!} \otimes \ket{1} \bra{n} + \hc \right] + \\
+ \sum_{n>1} \left[ R_{*n}(\vec{k}) e^{i \omega t} \ket{\! \downarrow} \bra{\downarrow \!} \otimes \ket{1} \bra{n} + \hc \right].
\label{eq:LatticeShakingHamBfields}
\end{multline}
The $\vec{k}$-dependent couplings are given by
\begin{flalign}
R_n(\vec{k}) &= \frac{1}{2} \left[ F_x \mathcal{A}_x^{1,n}(\vec{k}) + F_y e^{i \phi_F} \mathcal{A}_y^{1,n}(\vec{k}) \right] \\
R_{*n}(\vec{k}) &= \frac{1}{2} \left[ F_x \mathcal{A}_{*,x}^{1,n}(\vec{k}) + F_y e^{i \phi_F} \mathcal{A}_{*,y}^{1,n}(\vec{k}) \right],
\end{flalign}
where the Berry connection in the $\uparrow$ sector is
\begin{equation}
\vec{\mathcal{A}}^{n,m}(\vec{k}) = \bra{u_n(\vec{k})} i \nabla_{\vec{k}}  \ket{u_m(\vec{k})}
\end{equation}
and in the $\ket{ \! \downarrow}$ sector it is given by
\begin{multline}
\vec{\mathcal{A}}^{n,m}_*(\vec{k}) = \bra{K u_n(-\vec{k})} i \nabla_{\vec{k}} K \ket{u_m(-\vec{k})} \\
 = \l \vec{\mathcal{A}}^{n,m}(-\vec{k}) \r^* .
\label{eq:Astar}
\end{multline}

\emph{MW coupling.--} To obtain an effective SOC Hamiltonian for the two pseudospin states 
\begin{equation}
\ket{\! \Uparrow, \vec{k}} = \ket{\! \uparrow} \ket{u_1(\vec{k})}, \qquad  \ket{\! \Downarrow, \vec{k}} = \ket{\! \downarrow} K \ket{u_1(-\vec{k})},
\end{equation}
we introduce MW transitions $\Omega_{\rm MW}^{(1,2)}$ between $\ket{ \! \! \uparrow}$ and $\ket{ \! \! \downarrow}$ states with frequencies $\omega_{1,2}=\omega_0 \pm \omega$ as shown in Fig.~\ref{MagneticFieldSOC}. Assuming that $\omega_0 \gg \omega$ we obtain within the RWA
\begin{equation}
\H_{\rm MW}^{(j)}  \stackrel{\rm RWA}{\approx}  \frac{\Omega_{\rm MW}^{(j)}}{2} e^{i ( \omega_j t + \phi_{\rm MW}^{(j)} ) }  \ket{ \! \uparrow} \bra{\downarrow \!} + \hc
\end{equation}
Projection onto the band eigenstates gives rise to the following effective MW couplings 
\begin{equation}
\H_{\rm MW}^{(j)} = \sum_{n,m}  \frac{\Omega_{\rm MW}^{(j)}}{2} \l K^{m,n}(\vec{k}) \r^* e^{i ( \omega_j t + \phi_{\rm MW}^{(j)} ) }  \ket{ \! \uparrow,n} \bra{\downarrow, m} + \hc
\end{equation}
which are renormalized by Franck-Condon factors,
\begin{equation}
K^{m,n}(\vec{k}) = \langle K u_m(-\vec{k}) \ket{u_n(\vec{k})}.
\end{equation}

\emph{Symmetries.--} Before deriving the effective Hamiltonian, we make some symmetry considerations that will allow us to understand under which conditions the system is TR invariant. We will assume that the system is inversion symmetric and we make a gauge choice respecting this symmetry. Hence $\hat{P} \H_0(\vec{k}) \hat{P} = \H_0(-\vec{k})$ and Bloch states at $\pm \vec{k}$ are related by
\begin{equation}
\hat{P} \ket{u_{n}(\vec{k})} = e^{i \chi_n(\vec{k})} \ket{u_n(-\vec{k})},
\label{eq:PukProp}
\end{equation}
where $\chi_n(\vec{k})$ is a gauge-degree of freedom. It follows that 
\begin{equation}
\vec{\mathcal{A}}^{n,m}(-\vec{k}) = - e^{- i \left[ \chi_n(\vec{k}) - \chi_m(\vec{k}) \right]} \vec{\mathcal{A}}^{n,m}(\vec{k}),
\end{equation}
and from Eq. \eqref{eq:Astar} we derive that
\begin{equation}
\vec{\mathcal{A}}^{n,m}_*(\vec{k}) = - e^{- i \left[ \chi_n(\vec{k}) - \chi_m(\vec{k}) \right]} \vec{\mathcal{A}}^{m,n}(\vec{k}).
\end{equation}

For the lattice shaking matrix elements we obtain the relation
\begin{equation}
R_{*,n}(\vec{k},\phi_F) = - \left[ R_n(\vec{k},-\phi_F) \right]^* e^{- i \left[ \chi_1(\vec{k}) - \chi_n(\vec{k}) \right]}
\end{equation}
and it follows that
\begin{flalign}
|R_{n}(-\vec{k},\phi_F)|^2 &= |R_{n}(\vec{k},\phi_F)|^2,\\
|R_{*,n}(-\vec{k},\phi_F)|^2 &= |R_{n}(\vec{k},-\phi_F)|^2.
\end{flalign}
These expressions appear in the AC Stark shifts caused by lattice shaking. Unless $\phi_F = - \phi_F$ the amplitudes for $\ket{ \! \! \uparrow}$-states ($|R_n|^2$) and for $\ket{ \! \! \downarrow}$-states ($|R_{*,n}|^2$) differ, breaking TR invariance. Hence we will restrict ourselves to values
\begin{equation}
\phi_F = 0, \pi.
\label{eq:phiFforTR}
\end{equation}

For the matrix elements associated with MW transitions, we obtain from $\hat{P} K \hat{P} = K$ that
\begin{equation}
K^{n,m}(-\vec{k}) = e^{i \left[ \chi_n(\vec{k}) - \chi_m(\vec{k}) \right]} K^{n,m}(\vec{k}).
\end{equation}
Notably, the following product is an anti-symmetric function of $\vec{k}$ independent of the gauge choice $\chi_n(\vec{k})$,
\begin{equation}
R_n^*(\vec{k}) K^{1,n}(\vec{k}) = - R^*_n(-\vec{k}) K^{1,n}(-\vec{k}).
\label{eq:RKsymm}
\end{equation}
As will be shown next, this symmetry leads to a TR invariant effective Hamiltonian.

\emph{Effective SOC.--} 
From second order perturbation theory, in $F$ and $\Omega_{\rm MW}$, we derive the effective Hamiltonian in the subspace spanned by $\ket{\! \Uparrow}$ and $\ket{\! \Downarrow}$. As explained around Eq.\eqref{eq:phiFforTR} we consider the case when $\phi_F = - \phi_F$ and $\Omega_{\rm MW}^{(1)}=\Omega_{\rm MW}^{(2)}=\Omega_{\rm MW}$ to avoid TR breaking by spin-dependent AC Stark shifts from lattice shaking and MW transitions, respectively. As a result we obtain for $\phi_{\rm MW}^{(1)} = \phi_{\rm MW}+\pi$ and $\phi_{\rm MW}^{(2)} = \phi_{\rm MW}$,
\begin{multline}
\H_{\rm eff}(\vec{k}) = \epsilon_1(\vec{k}) + \sum_{n>1} \frac{\l \Omega_{\rm MW}\r^2 |K^{n,1}(\vec{k})|^2}{4 \Delta_n(\vec{k})} + \frac{|R_n(\vec{k})|^2}{\Delta_n(\vec{k})} \\
 +\sum_{n>1} \frac{\Omega_{\rm MW}}{\Delta_n(\vec{k})} \left[ \ket{\! \Downarrow} \bra{\Uparrow \!} e^{-i \phi_{\rm MW}}  R_n^*(\vec{k}) K^{1,n}(\vec{k})  + \hc \right].
 \label{eq:HeffSOCmagn}
\end{multline}
The detuning $\Delta_n(\vec{k}) = \epsilon_1(\vec{k}) - \epsilon_n(\vec{k}) + \omega$ is symmetric in $\vec{k}$, i.e. $\Delta_n(-\vec{k}) = \Delta_n(\vec{k})$.

Using the symmetry in Eq.\eqref{eq:RKsymm} it follows that the effective Hamiltonian in Eq.\eqref{eq:HeffSOCmagn} is TR invariant and represents an effective spin-orbit interaction. The first line of $\H_{\rm eff}(\vec{k})$ denotes the spin-independent dispersion relation, renormalized by AC Stark shifts. An alternative way to check TR invariance of the effective Hamiltonian is to study the original time-dependent Hamiltonian $H(t)$ (before applying RWA and perturbation theory). As shown in Ref.~\cite{Kitagawa2010} it is sufficient to proof that it is time-reversal invariant, $H(t_0 + t) = H(t_0 - t)$ for some $t_0$. We checked that this is the case for the situation discussed in this part of the supplementary.


\end{document}